\documentstyle[aps,prl,epsfig]{revtex}
\begin{document}
\draft
\twocolumn[\hsize\textwidth\columnwidth\hsize\csname %
@twocolumnfalse\endcsname

\title{Hall constant of strongly correlated electrons on a ladder}
\author{ P. Prelov\v sek$^{1,2}$, M. Long$^3$, T. Marke\v
z$^{2}$ and  X.Zotos$^4$ }
\address{$^1$ J. Stefan Institute, University of Ljubljana,
1001 Ljubljana, Slovenia }
\address{ $^2$ Faculty of Mathematics and Physics, University of
Ljubljana, 1000 Ljubljana, Slovenia }
\address{ $^3$ Department of Physics, University of Birmingham,
Edgbaston, Birmingham B15 2TT, United Kingdom}
\address{ $^4$ Institut Romand de Recherche Num\'erique en Physique des
Mat\'eriaux (EPFL-PPH), CH-1015 Lausanne, Switzerland }
\date{\today}
\maketitle
\begin{abstract}\widetext
The Hall constant $R_H$ in a tight-binding model of correlated
electrons on a ladder at $T=0$ is expressed in terms of derivatives of
the {\it ground state energy} with respect to external magnetic and electric
fields. This novel method is used for the analysis of the $t$-$J$ model on
finite size ladders.  It is found that for a single hole $R_H$ is
hole-like and close to the semiclassical value, while for two holes it
can vary with ladder geometry. In odd-leg ladders, $R_H$ behaves
quite regularly changing sign as a function of doping, the variation being
quantitatively close to experimental results in cuprates.

\end{abstract}
]
\narrowtext
The Hall response in materials with strongly correlated electrons
remains one of the properties least understood theoretically. The
subject has been stimulated by experiments in superconducting cuprates
\cite{ong}, revealing anomalous doping and temperature dependence of
the Hall constant $R_H(T)$ in the normal metallic state. For instance,
it is well established that the Hall effect is hole-like, $R_H>0$, in
materials with a low density of holes $n_h$, introduced by doping the
reference antiferromagnetic (AFM) insulator.  The clearest realization
is $La_{2-x}Sr_xCuO_4$ (LSCO), where the doping $x$ can be directly
related to the concentration of mobile holes per unit cell $n_h=x$ and
the semiclassical result $R_H= 1/n_h e_0$ seems to be obeyed at lowest
$T>T_c$ and at low doping \cite{ong,hwan}.

Theoretical attempts to calculate the Hall effect in models of
strongly correlated electrons resulted in quite controversial
conclusions. Even for weak correlations \cite{fuku} or for the problem
of a single carrier in a Mott-Hubbard insulator \cite{brin}, the
analysis of the Hall response is fairly involved. In more recent
investigations relevant to cuprates, the dynamical Hall constant
$\tilde R_H(\omega)$ has been studied within linear response theory
for the $t$-$J$ and Hubbard model, analytically by high-$\omega,T$
expansion \cite{shas} and numerically via exact-diagonalization
studies of small systems \cite{cast,assa}.  The obtained results are
quite consistent for the high-frequency quantity $R_H^*=\tilde
R_H(\omega \to \infty)$, showing at high $T$ a transition from a
hole-like, $R_H^*>0$, to an electron-like, $R_H^*<0$, at a finite
crossover $n_h^*\sim 1/3$ \cite{shas}.  For the most interesting
d.c. limit $R_H=\tilde R_H(\omega=0)$ the majority of results obtained
for 2D systems at low doping and $T\to 0$ indicate $R_H<0$
\cite{assa}, instead of the expected hole-like behavior \cite{rojo}.
On the other hand, one of the present authors \cite{prel} recently
showed that for a single hole doped into a 2D AFM at $T=0$ the result
should be the semiclassical one with $R_H>0$.

>From another perspective and stimulated by synthesis and experiments on
novel cuprates, models of interacting electrons on ladder systems have
also been extensively studied in recent years \cite{dago}.  The idea
is that ladders with a variable number of legs can offer a broader
insight into the behavior of correlated electrons and thus can lead to
an understanding of the more challenging 2D systems.  Again, results
for the Hall response $\tilde R_H(\omega)$ at low doping obtained
through linear response theory reveal a quite puzzling, electron-like,
$R_H<0$ \cite{tsun} behavior.

Our aim in this work is to formulate and calculate the Hall constant
as a ground state ($T=0$) property for a tight-binding model with a
ladder geometry.  This is possible due to the finite transverse width
of the system that, in contrast to an infinite 2D (or higher D)
system, does not require a relaxation mechanism to describe a proper
transport regime.  Such a formulation allows for a more transparent
calculation of $R_H$ and in particular the determination of its sign
\cite{rojo}.  In the following, we apply this method to the $t$-$J$
model on a ladder.  Via a numerical analysis of small systems, we
investigate $R_H$ for few holes $N_h=1,2$ and a finite concentration
of holes $n_h>0$ introduced into an AFM correlated spin background.

Let us consider the simplest single-band tight-binding model of
interacting fermions on a ladder geometry with $M$ legs in the $y$
direction, $L$ rungs in the $x$ direction. Periodic boundary
conditions (p.b.c.) are assumed in the $x$ direction and unit-cell
length $a_0=1$.  To analyze the Hall response, the following
additional ingredients need to be incorporated into the model:

\noindent a) a finite transverse electric field ${\cal E}_y= {\cal E}
\neq 0$ has to be taken into account,

\noindent b) a homogeneous magnetic field $B$ perpendicular to the
ladder, introduced via the Peierls substitution.  In the Landau gauge
${\bf A}=B(-y, 0)$, only the phases of hopping integrals in the
$x$ direction, $H_x$, are
modified by a phase $\varphi = e_0 B$ ($\hbar=1$). We also assume that
the interaction term $H_{int}$ is not influenced neither by ${\bf A}$
nor by ${\cal E}$.

\noindent c) a steady electric current density $j=j_x$
is induced in the ground state by piercing a closed ladder
in the $y$ direction with a flux $\Phi$, modifying
the hopping term by a phase $\theta= e_0 \Phi/L$.

The tight-binding model can be thus written as,
\begin{eqnarray}
H&=&H_x+H_y+H_{\Delta}+H_{int} \nonumber \\
H_x &=&-t \sum_{m=1}^M \sum_{is} {\rm e}^{i[(m-\bar m) \varphi
-\theta ]} (c^{\dagger}_{m,i+1,s} c_{mis} + H.c.),
\nonumber \\ H_y&=& -t'
 \sum_{m=1}^{M-1} \sum_{is} (c^{\dagger}_{m+1,is} c_{mis} + H.c.),
\label{eq1} \\
H_{\Delta}&=& \Delta \sum_{im} (m-\bar m) n_{mi},\nonumber
\end{eqnarray}
where $\Delta =e_0 {\cal E}$, $\bar m=(M+1)/2$ and $H_{int}$ will be
chosen later on.

The idea is to study the ground state energy
$E(\theta,\Delta,\varphi)$ of the system as a function of
$\Delta, \theta,\varphi$ in order to evaluate the Hall constant $R_H$,
\begin{equation}
R_H = -\frac{{\cal E}}{j B} = -\frac{\Delta}{j \varphi}.
\label{eq3}
\end{equation}
We use the fact that the electric current density $j$
and polarization density $P=P_y$ in the ground state
$|0\rangle$ can be evaluated via derivatives of the
energy $E(\theta,\Delta,\varphi)$ using the Feynman-Hellmann relations,
\begin{equation}
j = \frac{e_0}{N}\frac{\partial E}{\partial\theta},\qquad
P = -\frac{e_0}{N}\frac{\partial E}{\partial\Delta},
\label{eq4}
\end{equation}
where $N=LM$ denotes the number of lattice points.

In the absence of the magnetic field, $\varphi=0$, we use as starting
point an equilibrium and nonpolar ground state with $j=0$ and
$P=0$. Such a ground state might correspond to finite values
$\theta_0$ and $\Delta_0$. In ladders one expects (for a nondegenerate
ground state) $\Delta_0=0$ by symmetry. On the other hand, in finite
systems, in general we find $\theta_0 \neq 0$. This can be considered
as a finite size effect, since in the thermodynamic limit $L\to
\infty$ no macroscopic current is expected in the ground state and so
$\theta_0\to 0$ (for a particular study of finite-size scaling of
$\theta_0$ see e.g. Ref.\cite{zoto}). Taking a proper starting
$\theta_0$ in the following calculations is however crucial for
obtaining a sensible result.

Next, to simulate the Hall effect we analyze systems with small but
finite current $j\neq 0$ imposed by a finite $\tilde\theta$, magnetic
field imposed by a $\varphi\neq 0$ and choosing a $\Delta\neq 0$ so
that the system remains nonpolar, $P=0$.  Hence we have to study the
variation of $E(\theta_0+\tilde\theta,\Delta,\varphi)$ for small
$\tilde\theta,\Delta,\varphi$. It is enough to consider a Taylor
expansion up to 3$^{rd}$ order, simplified by invoking: (i) the
symmetry of the current operator $\hat j_0
=\frac{e_0}{N}\frac{\partial H}{\partial \theta}
(\tilde\theta\neq0,\varphi=0)$ under reflection $(\Delta\rightarrow
-\Delta)$ and, (ii) the reflection antisymmetry of the diamagnetic current
$\hat
j_0^a =\frac{e_0}{N}\frac{\partial H}{\partial\varphi}
(\tilde\theta\neq0,\varphi=0)$ and the polarization operator $\hat P
=-\frac{e_0}{N}\frac{\partial H}{\partial \Delta}
(\tilde\theta=0,\varphi=0)$. This leads to:
\begin{eqnarray}
E=&&E^0+\frac{1}{2}E^0_{\theta\theta}\tilde \theta^2 +
\frac{1}{2}E^0_{\Delta\Delta} \Delta^2 + E^0_{\Delta\varphi}
\Delta\varphi + \nonumber \\
&+&E^0_{\theta\Delta\varphi}\tilde\theta\Delta\varphi
+\frac{1}{2} E^0_{\theta\Delta\Delta}\tilde\theta\Delta^2
+\frac{1}{6}
E^0_{\theta\theta\theta} \tilde\theta^3 +\cdots.
\label{eq10}
\end{eqnarray}
The superscript zero indicates derivatives at equilibrium (at
$\varphi=0$).  From Eqs.(\ref{eq4},\ref{eq10}) and to leading order,
$j$ is given by
\begin{equation}
j=\frac{e_0}{N} E^0_{\theta\theta} \tilde\theta, \label{eq11}
\end{equation}
while the Hall field $\Delta$ is set by the condition that
$P=0$ even in the presence of finite $\tilde \theta,\varphi$,
\begin{equation}
E^0_{\Delta\Delta}\Delta+E^0_{\Delta\varphi}\varphi +
E^0_{\theta\Delta\varphi}\tilde\theta\varphi+E^0_{\theta\Delta\Delta}\tilde
\theta\Delta =0. \label{eq12}
\end{equation}
Retaining terms linear in $\varphi$ and $\tilde \theta$ we
can express $\Delta$ as
\begin{eqnarray}
&\Delta&=\Delta_{\varphi}+\Delta_j=
-\frac{E^0_{\Delta\varphi}}{E^0_{\Delta\Delta}}\varphi
-\frac{\tilde E^0_{\theta\Delta\varphi}}{E^0_{\Delta\Delta}}
\varphi\tilde\theta, \nonumber \\
&\tilde E^0_{\theta\Delta\varphi}&= E^0_{\theta\Delta\varphi}-
\frac{E^0_{\Delta\varphi}E^0_{\theta\Delta\Delta}}{E^0_{\Delta\Delta}}.
\label{eq13}
\end{eqnarray}
We are interested in the second term, i.e. in $\Delta_j$ induced by
finite $\tilde \theta$ and related $j$. We note also that $\tilde
E^0_{\theta\Delta\varphi}$ can be expressed simply as the derivative
taken at the origin $\Delta_0=0$ shifted to $\Delta_{\varphi}$.
Inserting $\Delta_j$ from Eq.(\ref{eq13}) and $j$ from Eq.(\ref{eq11})
into Eq.(\ref{eq3}), we obtain the expression
\begin{equation}
R_H= \frac{ N \tilde E^0_{\theta\Delta\varphi}}{e_0 E^0_{\Delta
\Delta} E^0_{\theta \theta}}. \label{eq15}
\end{equation}
This is a central result in this work.
The main advantages of the new approach are:
(i) Eq.(\ref{eq15}) requires the knowledge of only the ground
state energy, (ii) the condition for $j=0,~P=0$ in the reference ground
state of finite size systems is much more transparent.

Now we will show how this formulation is related to the linear
response theory for the particular case of ladders at $T=0$, where
$R_H$ is evaluated via the dynamic (in general complex) $\tilde
R_H(\omega)$ \cite{shas},
\begin{eqnarray}
\tilde R_H(\omega)&=& -\frac{1}{B}
\frac{\sigma_{yx}(\omega)}{\sigma^0_{xx}(\omega)
\sigma^0_{yy}(\omega)}, \nonumber \\
\sigma_{\alpha\beta}(\omega)&=& \frac{i e_0^2}{\omega N} \bigl(
\langle\tau_{\alpha\beta} \rangle -
\frac{N^2}{e_0^2}\int_0^{\infty} dt {\rm e}^{i\omega t}
\langle [\hat j_{\alpha}(t),\hat j_{\beta}]\rangle).
\label{eq17}
\end{eqnarray}
$\sigma_{\alpha\beta}$ denotes the conductivity tensor evaluated
at $B \neq 0$, while $\sigma^0_{\alpha\beta}$ at $B=0$.
$\langle\tau_{\alpha\beta} \rangle$ are stress (kinetic energy)
tensor components, in ladders nonvanishing only in the direction of
p.b.c., i.e. for $\alpha=\beta=x$.

We are interested in the limit $\omega \to 0$.
At $T=0$ and with p.b.c. in the $x$ direction,
$\sigma^0_{xx}$ describes a dissipationless singular conductivity given by
\begin{equation}
\sigma^0_{xx}(\omega \to 0)= \frac{2 i e_0^2}{\omega} D_{xx}
=\frac{i e_0^2}{\omega}E^0_{\theta\theta},
\label{eq18a}
\end{equation}
where $D_{xx}$ denotes the charge stiffness \cite{kohn,zoto}.
On the other hand, in the $y$ direction
the polarizability $\chi_{yy}(\omega \to 0)$ is finite due to open
boundaries. Using the relation $\hat j_y=d \hat P/dt$, we get
from Eq.(\ref{eq17})
\begin{equation}
\sigma^0_{yy}(\omega \to 0)=i \omega \chi_{yy}
=-\frac{i \omega e_0^2}{N}E^0_{\Delta\Delta}.
\label{eq18b}
\end{equation}
The off-diagonal $\sigma_{yx}(\omega=0)$ can be written as
\begin{eqnarray}
\sigma_{yx}(0)=-i N\int_0^{\infty} dt \langle [\hat P(t),\hat
j]\rangle &=& \nonumber \\
= 2 N\sum_m \frac{ \langle 0|\hat P|m\rangle
\langle m |\hat j|0\rangle} {E_0-E_m}
= -e_0 \frac{\partial \langle 0|\hat j|0 \rangle}{\partial \Delta} &=&
-\frac{e_0^2 E_{\theta\Delta}}{N},
\label{eq20}
\end{eqnarray}
where the ground state $|0\rangle$, excited states $|m\rangle$ as well
as $E_{\theta\Delta}$ refer to $\varphi \neq 0$ but $\Delta=0$.
Taking into account that $E_{\theta\Delta} \propto \varphi$ and
inserting relations (\ref{eq18a}-\ref{eq20}) into Eq.(\ref{eq17}) we
recover the expression (\ref{eq15}), provided that
$E^0_{\Delta\varphi}=0$. The equivalence for the case
$E^0_{\Delta\varphi}\neq 0$ can also be obtained if one calculates the
linear response $\sigma_{yx}$ in Eq.(\ref{eq20}) not at $\Delta=0$,
but rather at the proper $\Delta=\Delta_{\varphi}$.

>From linear response theory we observe that the existence of the
simple expression Eq.(\ref{eq15}) is subject to the presence of the
restricted geometry which implies a finite $\sigma_{yx}(\omega=0)$ as
well as a finite $\sigma_{xx}(\omega)\sigma_{yy}(\omega)$ for
$\omega\to 0$.  Both quantities would diverge for $T=0$ at the 2D
limit ($M\to \infty$), although $R_H(\omega \to 0)$ is expected to
remain well defined and bounded \cite{prel}.

Let us first test the method for noninteracting electrons on a two-leg
($M=2$) ladder. It is here easy to find the single-electron
eigenenergies $\epsilon_{\pm}(k,\theta,\Delta,\varphi)$, referring to
the upper and lower band. At $T=0$, states in both bands are
occupied for $\epsilon^{\pm}<\epsilon_F$, and the result follows from
Eq.(\ref{eq15}),
\begin{equation}
R_H = \frac{\tau^+ -\tau^-}{e_0(\tau^- +\tau^+)(n_e^-
-n_e^+)}, \label{eq24}
\end{equation}
where $n_e^{\pm}$ are electron densities in both bands and
$\tau^{\pm}= 4 t \sum_{|k|<k^{\pm}} \cos k$. Note that
Eq. (\ref{eq24}) reduces to plausible expressions: a) the
semiclassical result $R_H=-1/n_e e_0$ for an empty upper band, $n_e^+=0$,
and b) $R_H=1/n_he_0$ for a filled lower band $n^-_e=1$,
where $n_h=1-n_e^+=2-n_e$ is the density of holes in the upper band.

Now, we illustrate the method and expected as well as anomalous
features of the Hall response in correlated systems on a study of the
isotropic $t$-$J$ model ($t'=t$). The interaction term in
Hamiltonian (\ref{eq1}) describes AFM exchange interaction
between fermionic spins on neighboring sites $j=(m,i)$,
\begin{equation}
H_{int}= J \sum_{\langle jj'\rangle } \vec S_j \cdot \vec S_{j'},
\label{eq25}
\end{equation}
and fermionic operators in the kinetic energy term are replaced by
projected ones, forbidding a double occupancy of sites. Note that the
projection does not influence the general formalism,
Eqs.(\ref{eq3}-\ref{eq15}).

To obtain $E(\theta,\Delta,\varphi)$ and consequently $R_H$ in finite
size ladders we employ the Lanczos diagonalization technique.
For a particular system with given $M,L$ and fixed number of holes
$N_h$ we first find the energy minimum at the equilibrium $\theta_0$
and then calculate numerically at this point derivatives
$E^0_{\Delta\Delta}, E^0_{\theta\theta}, E^0_{\Delta\varphi},
E^0_{\theta\Delta\varphi}, E^0_{\theta\Delta\Delta}$. Then $R_H$ is
evaluated using relations (\ref{eq13},\ref{eq15}).

In Fig.~1a,b we show results for the dimensionless $r_H=e_0 R_H/N$ in
the case of a single hole $N_h=1$ on $M=2$ and $M=3,4$ ladders,
respectively, of varying length and as a function of $J/t$.  Note that
the semiclassical result in this case would be $r_H=1$. We should note
that in general we find here $\theta_0\neq 0,\pi$. Moreover the
influence of $\Delta_{\varphi}\neq 0$ is essential since $\tilde
E^0_{\theta\Delta\varphi}$ differs from $E^0_{\theta\Delta\varphi}$
significantly. E.g., for $M=2$, in the most relevant regime $J<t$ both
quantities can even be of a different sign.  We notice that results
for different $L$'s are quite consistent. $J=0$ is a special case with
a ferromagnetically polarized ground state, $S^{tot}=(N-1)/2$ and
hence $r_H=1$.  Also, as Fig.~1 shows, for $J>0$ $R_H$ is hole-like
with $r_H \agt 1$.  This means that the behavior is very close to the
semiclassical one \cite{prel}, but the deviation from the latter is finite
(although smaller for larger $M$) and seems to persist also for $L\to
\infty$.

\begin{figure}
\begin{center}
\epsfig{file=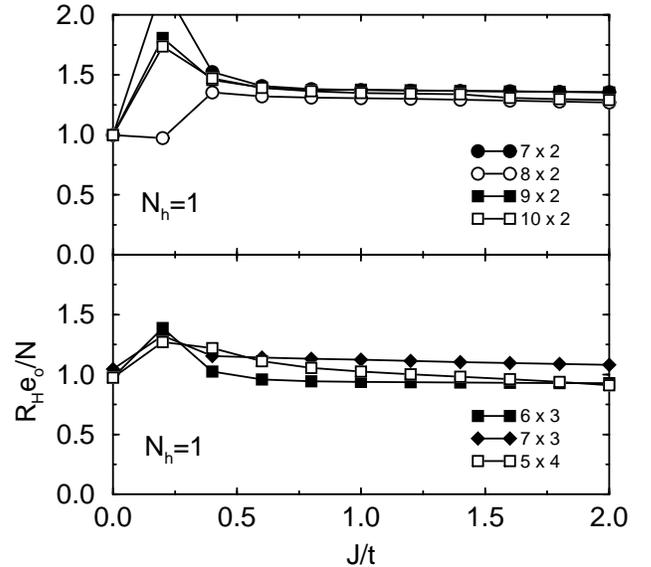,height=7.5cm}
\end{center}
\caption{Dimensionless Hall constant $r_H=e_0 R_H/N$ vs. $J/t$ for a
single hole on ladders of different lengths $L$ with: a) two legs and,
b) three and four legs.}
\label{fig1}
\end{figure}

Results for two holes are presented in Fig.~2. At very low doping, $n_h
\ll 1$, one might expect that in the thermodynamic limit holes behave
as independent particles so that $r_H \sim 1/N_h$.
This is definitely not the case for $M=2$, where in the
majority of the parameter regime we even find $r_H<0$, consistent with
Ref.\cite{tsun}. It seems that this phenomenon is related to the
existence of a spin gap in $M=2$ ladders and quite pronounced binding
of holes into pairs \cite{dago}.  Results appear more regular for
$M>2$.  Conclusions quite consistent with the semiclassical $r_H \sim
1/2$ are obtained for the $M=3$ and $M=4$ ladders.  It is well known that
odd-leg ladders do not show a spin gap \cite{dago}, so this can serve
as an explanation for the essential difference between the $M=3$ and
$M=2$ case.  For $M=4$, a small spin gap is expected in undoped
ladder \cite{dago}, however its effect on $r_H$ is not visible, at
least not for reachable $L$.

\begin{figure}
\begin{center}
\epsfig{file=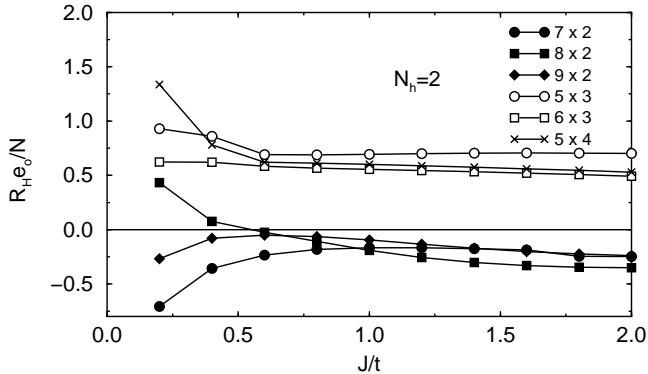,height=8.5cm,angle=-90}
\end{center}
\caption{$r_H=e_0 R_H/N$ vs. $J/t$ for two holes on ladders of
different size $L \times M$.}
\label{fig2}
\end{figure}

In systems with more holes, $N_h>2$, and available $N \leq 20$ we are
dealing already with a substantial doping $n_h$.  In Fig.~3 we show
results for the doping dependence $R_H(n_h)$.  We concentrate on a
more regular three-leg ladder, where even-odd effects in $N_h$
are not pronounced.  Shown are data for two systems, $5\times 3$ and
$6\times 3$. Results for both systems are in general quite consistent,
with deviations appearing only in the regime $n_h \sim 0.2$ where in
particular the value for $N_h=4$ on a $6\times 3$ system appears to be
irregular, probably a finite-size effect related e.g. to a
close-shell configuration.

\begin{figure}
\begin{center}
\epsfig{file=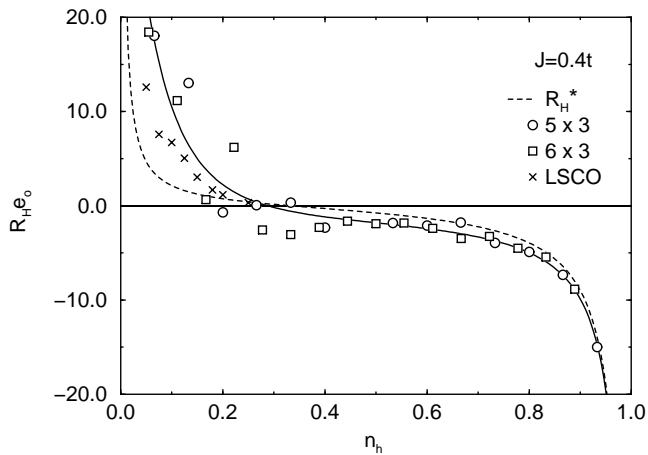,height=8.5cm,angle=-90}
\end{center}
\caption{Hall constant $e_0 R_H$ vs. hole doping $n_h$ for the
three-leg ladder and $J=0.4~t$. Results are shown for $L=5$ and $L=6$
systems, where the full line serves as a guide to the eye. Dashed line
represents the $R_H^*$ result [5]. Shown are also experimental
results for LSCO taken from Ref. [2]. }
\label{fig3}
\end{figure}

The interpretation of Fig.~3 is straightforward in two regimes. For a
nearly empty band $n_h \alt 1,~ n_e=1-n_h \agt 0$ we recover the
semiclassical result, $e_0R_H=-1/n_e$. Analogous, but only approximate,
is the hole-like behavior for low doping $n_h \ll 1$ where $e_0R_H
\sim 1/n_h$. The behavior is very asymmetric
between the hole and the electron side. The change from a hole-like
$R_H>0$ to an electron-like $R_H<0$ appears (with the largest
scattering of results in this regime) at $n_h^* \sim 0.27$.  Our value
for the crossover $n_h^*$ is close to the crossover in
$R_H^*=1/4n_h-1/(1-n_h)+3/4$ (in 2D and $T\to \infty$) at $n_h^* \sim
1/3$ \cite{shas}. In spite of similar values for $n_h^*$ and a
quantitative agreement for $n_h>0.3$, $R_H^*$ (also plotted in Fig.~3)
deviates at low doping values by factor of 4 from the semiclassical
result.

Although we are dealing with a ladder system, we expect that results
for odd-leg ladders would be very analogous to 2D systems. It is
therefore not surprising that our results for $R_H(n_h)$ are both
qualitatively as well quantitatively close to experimental ones for
LSCO (doping range $0<x<0.35$), where values
shown in Fig.~3 are taken at $T=100~K$ \cite{hwan}. We note that
experimentally the crossover appears at $x \sim 0.3$ and at low doping,
data are consistent with the semiclassical $R_H \sim 1/n_h e_0$.

In conclusion, we have introduced a novel method which allows the
evaluation of the d.c. Hall constant at $T=0$ in correlated systems
with a ladder geometry solely from the ground state energy.
Since the behavior of ladder systems is in
many respects analogous to 2D systems the method can be used to
approach the anomalous and theoretically controvertial $R_H$ in
cuprates. Our numerical results emerging from odd-leg ladders are
indeed surprisingly close to experiments in LSCO.

Part of this work was done during visits of (PP) and (ML) at IRRMA as
academic guests of EPFL.
(XZ) acknowledges support by the Swiss National Foundation
grant No. 20-49486.96, the University of Fribourg and the University of
Neuch\^atel.

\end{document}